\documentclass[11pt,twoside]{article}
\usepackage{asp2004}
\usepackage{psfig}
\usepackage{epsf}
\usepackage{graphicx}
\usepackage{lscape}

\begin{document}
\title{Classical nova explosions}
\author{Margarita Hernanz}
\affil{Institut d'Estudis Espacials de Catalunya (CSIC). Campus
Universitat Aut\`onoma de Barcelona,
Facultat de Ci\`encies, Torre C5-parell, 2$^a$ planta.
E-08193 Bellaterra (Barcelona), Spain}

\begin{abstract}
A review of the present status of nova modeling is made, with a special
emphasis on some specific aspects. What are the main nucleosynthetic products
of the explosion and how do they depend on the white dwarf properties
(e.g. mass, chemical composition: CO or ONe)? What's the imprint of
nova nucleosynthesis on meteoritic presolar grains?
How can gamma rays, if observed with
present or future instruments onboard satellites, constrain nova models
through their nucleosynthesis?  What have we learned about the turnoff of
classical novae from observation with past and present X-ray observatories?
And last but not least, what are the most critical issues concerning nova
modeling (e.g. ejected masses, mixing mechanism between core and
envelope)?
\end{abstract}

\section{Introduction}
Classical novae are explosions occurring on top of white dwarfs in close
binary systems of the cataclysmic variable type. Transfer of hydrogen-rich
matter from the main sequence star to the white dwarf is the driver of
the explosion, provided that the white dwarf is massive enough and that
the mass accretion rate is sufficiently low. In these conditions, matter
accumulates on top of the white dwarf until it reaches hydrogen ignition
conditions, with a pressure such that matter is degenerate. This leads to
a thermonuclear runaway because self adjustment of the envelope once
nuclear burning starts is not possible.

The general scenario for nova explosions thus seems quite well understood,
but there are still many open issues concerning the initial conditions
leading to the explosion (e.g. need of mixing between core and envelope
material) and some observed properties of the outburst (e.g. amount of
mass ejected or velocity distribution in the ejecta).

In this paper some of the basic observational properties of novae
are first reviewed.
Then the general scenario for nova explosions based on hydrogen
thermonuclear runaway is outlined, paying special attention to the
influence of the properties of the underlying white dwarf (e.g. mass
and chemical composition). The relevance of
nucleosynthesis in classical novae (e.g. chemical evolution of the Galaxy,
understanding of isotopic ratios in some meteoritic presolar grains) is
then analyzed, as well as the potential of gamma-rays to reveal the nuclear
processes taking place during nova outbursts. A snapshot of
the link between the nova outburst and its cataclysmic variable {\it host}
as revealed by recent X-ray observations of novae is finally
presented, followed by a (partial) list of the still open issues in nova
modeling.

\section{Basic observational properties of novae}
\begin{figure}[t]
\plottwo{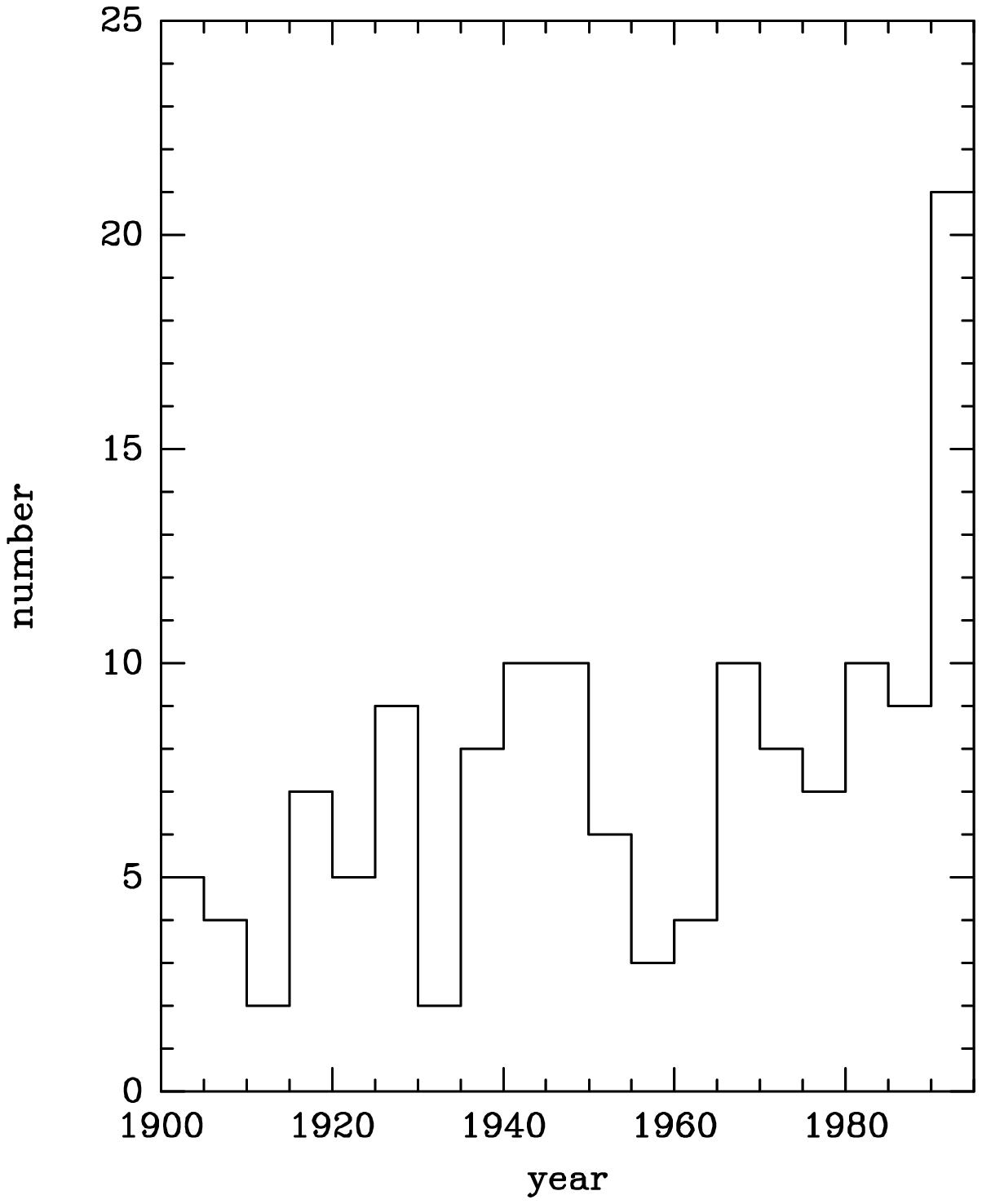}{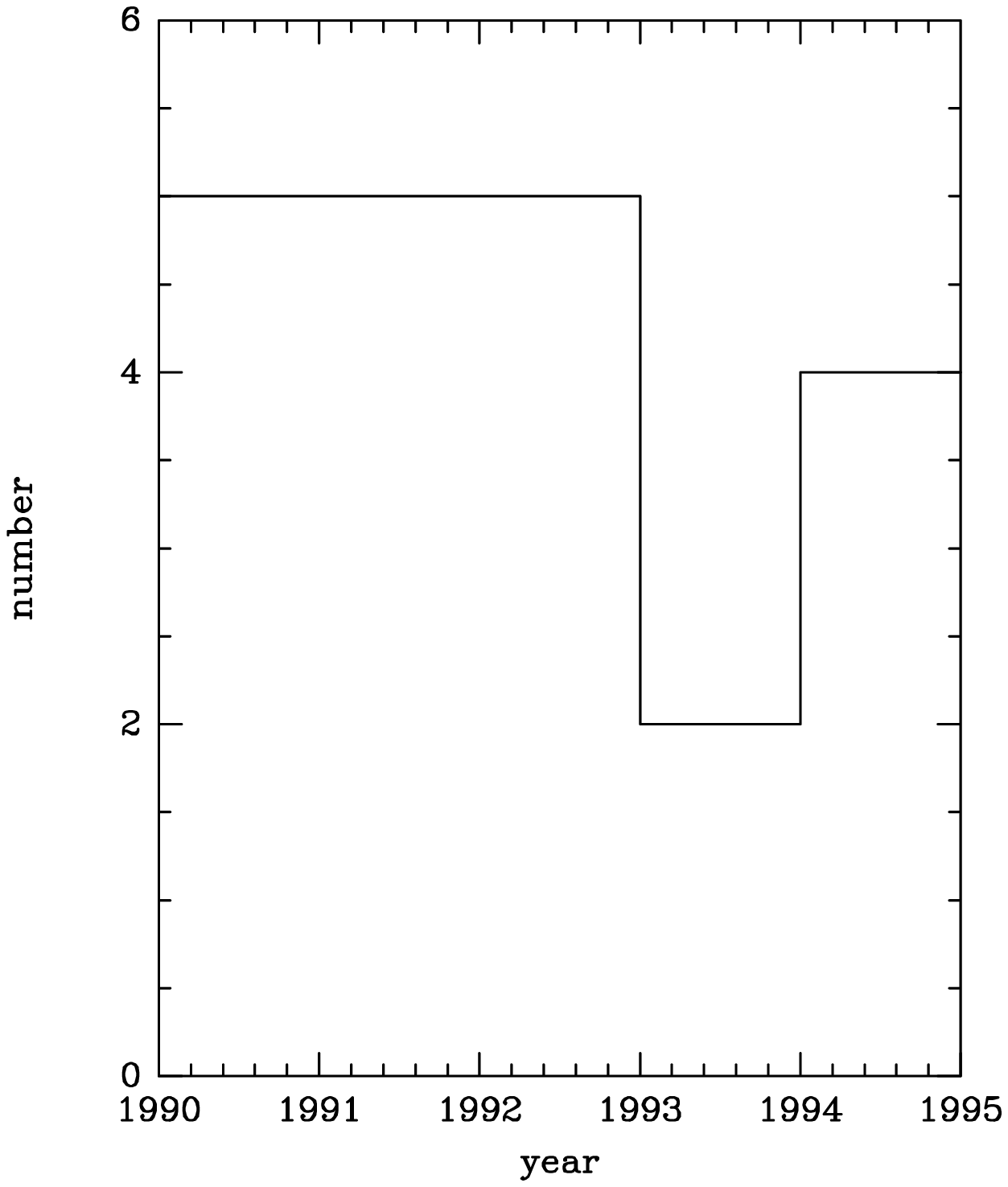}
\caption{Number of novae discovered during the last century. Left
panel: period 1901-1995. Right panel: period 1991-1995.}
\label{fig:discov}
\end{figure}

\begin{figure}[t]
\plottwo{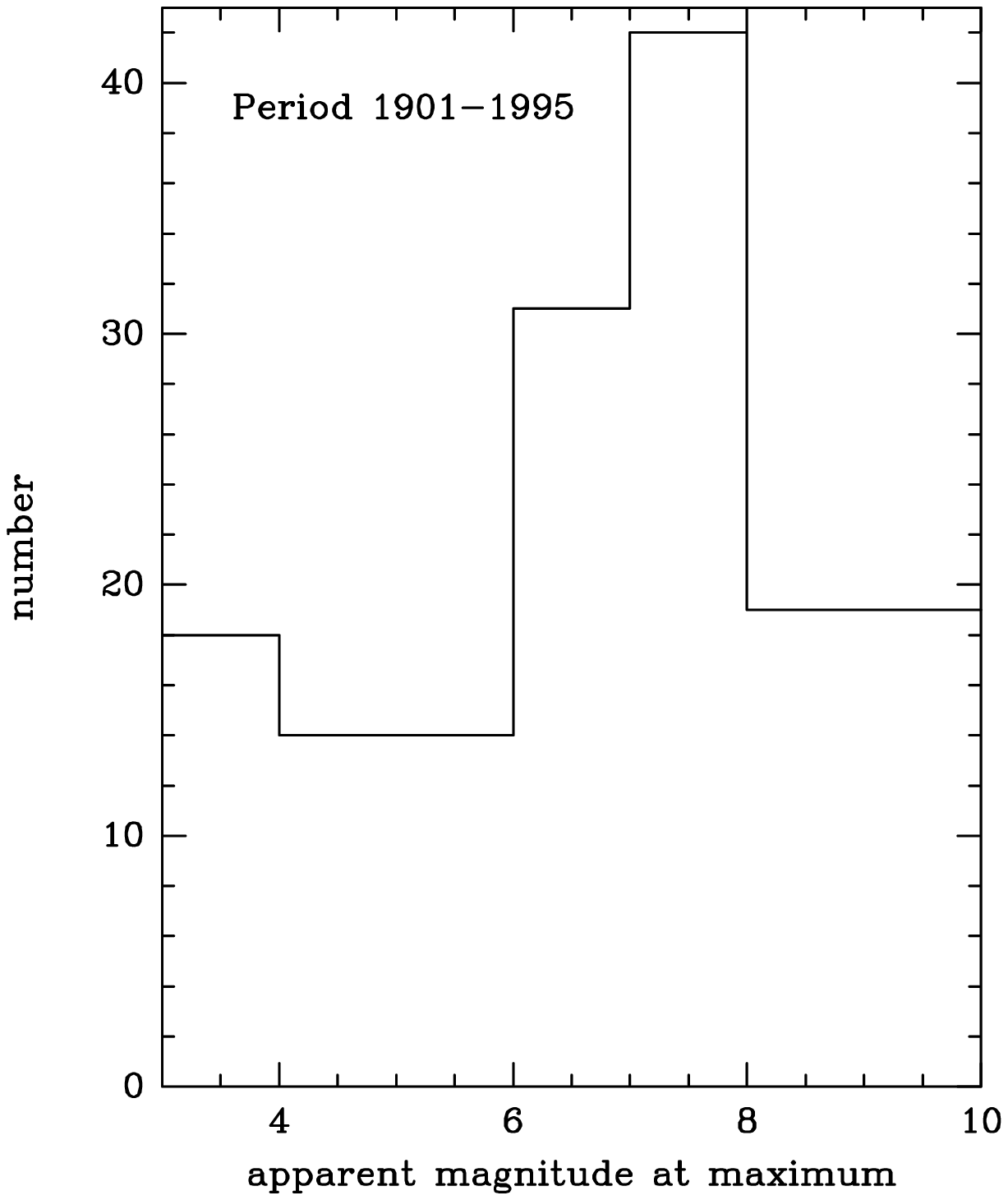}{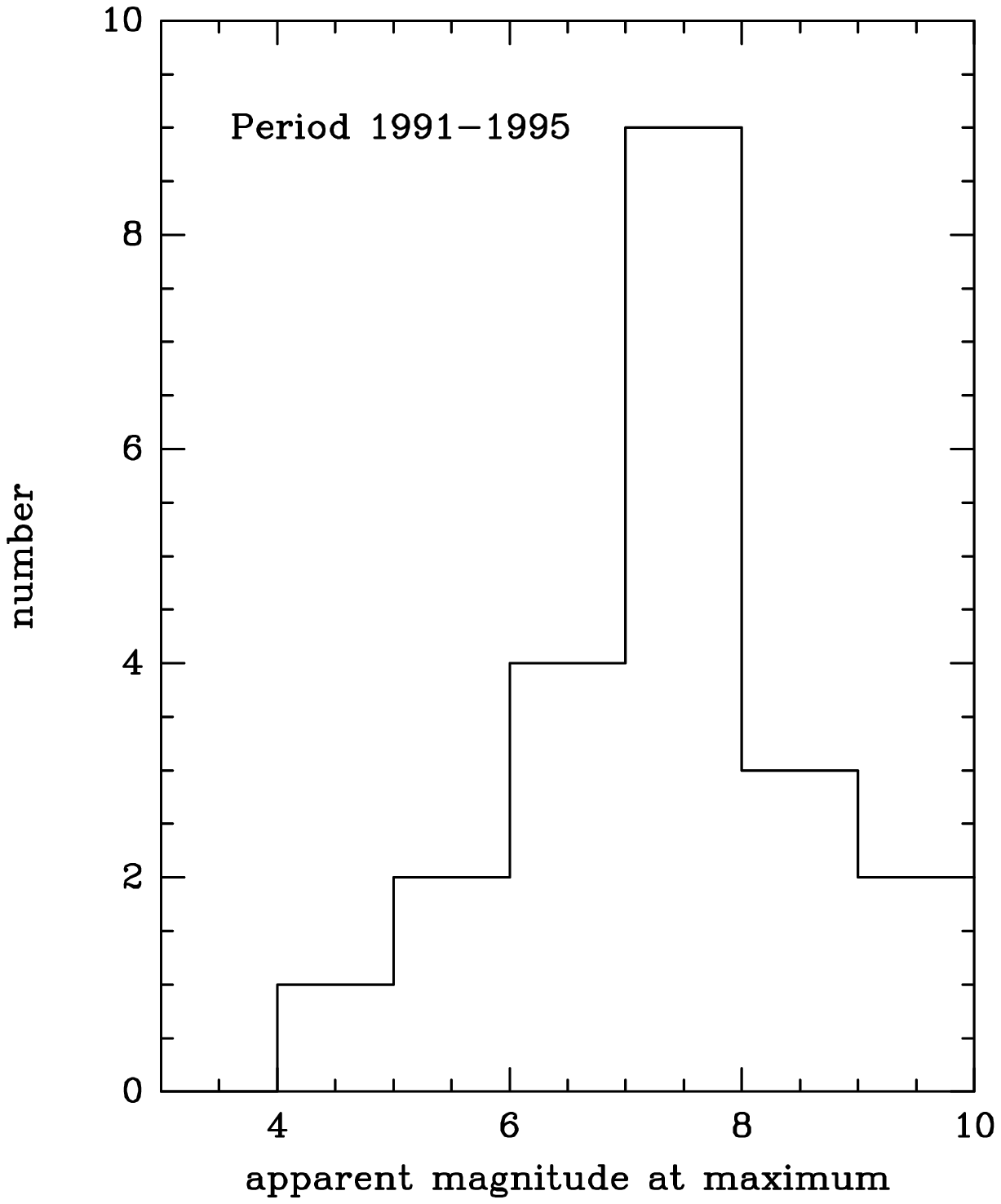}
\caption{Histogram of nova discoveries as a function of the visual
apparent magnitude at maximum. Left panel: period 1901-1995. Right
panel: period 1991-1995} \label{fig:histomv}
\end{figure}

\begin{figure}[t]
\plottwo{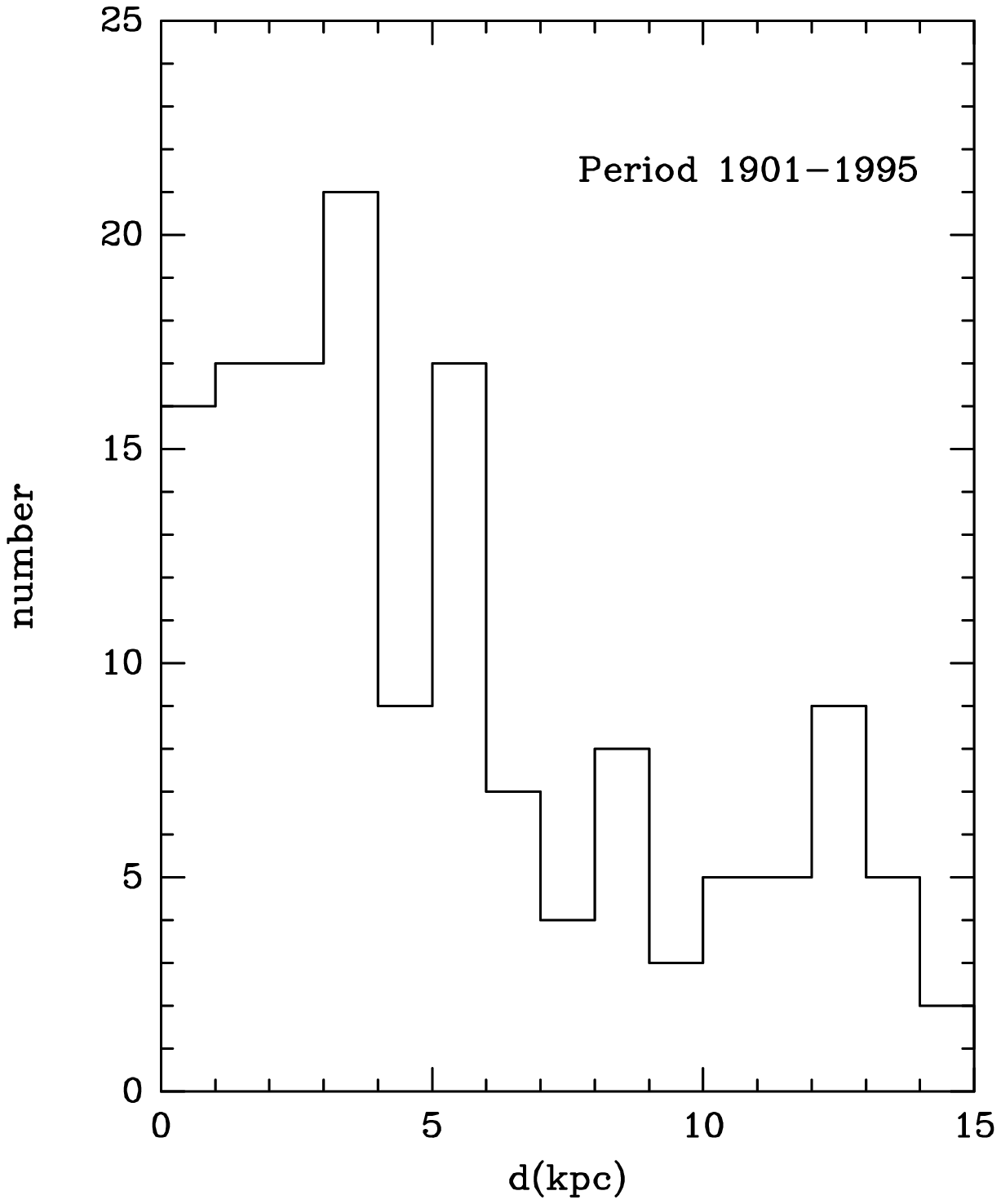}{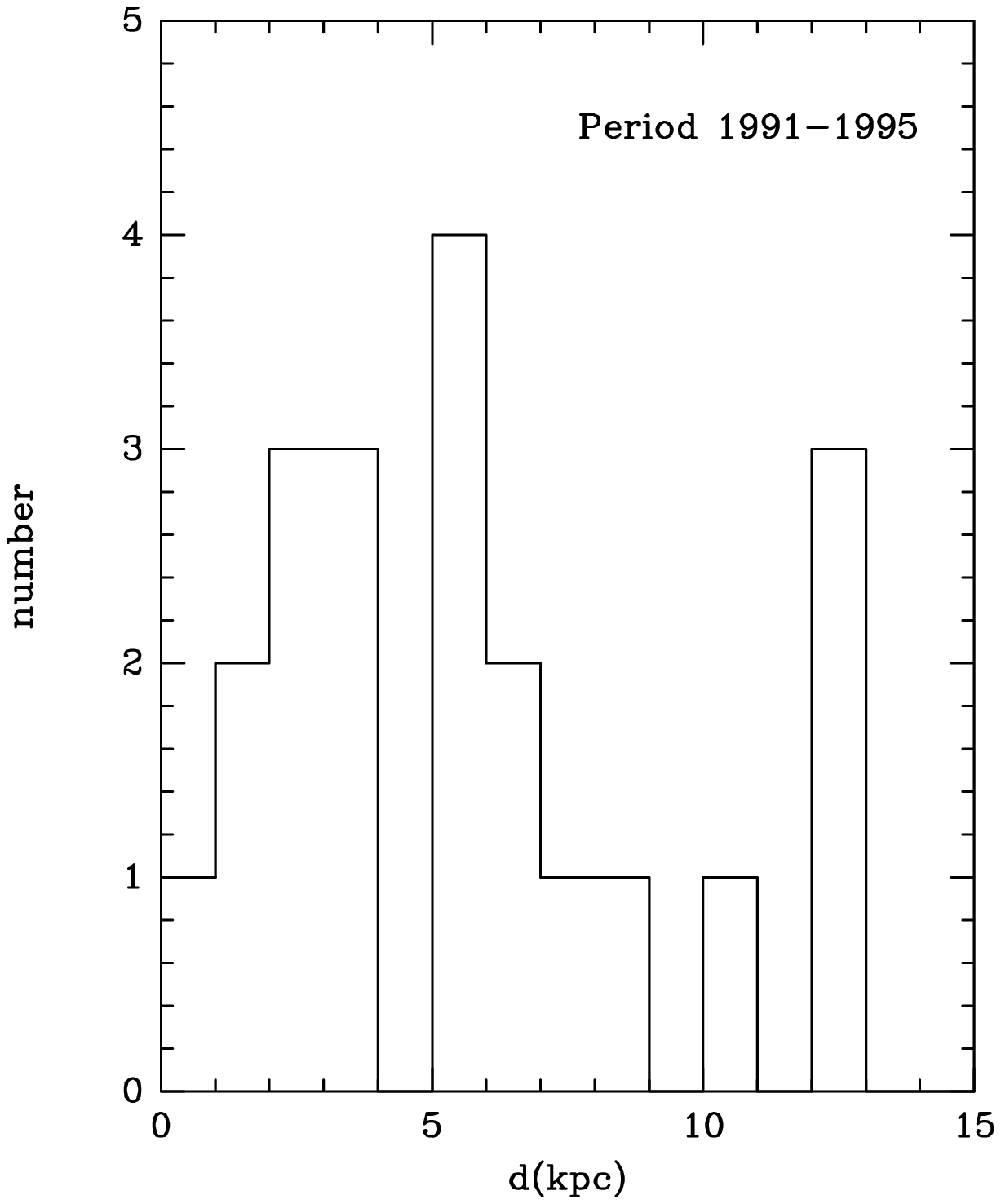}
\caption{Histogram of nova discoveries as a function of the
distance. Left panel: period 1901-1995. Right panel: period
1991-1995. Distances have been computed according to the MMRD in
\cite{Sha97}} \label{fig:histod}
\end{figure}

Classical novae are in general discovered by amateur astronomers,
looking at the sky to search for variabilities in stars. An
histogram of the number of nova discoveries in the last century
(period 1900-1995) is displayed in figure \ref{fig:discov}, based
on data from \cite{Sha97}, with a zoom of the most recent data,
1991-1995 (histograms versus apparent visual magnitude at maximum
and distance are also shown in figures \ref{fig:histomv} and
\ref{fig:histod}). Even if the dataset is not absolutely complete,
it can be seen that at most 5 novae are discovered every year in
our Galaxy. This number is far from the total number of novae
estimated to explode in the Milky Way, based either on
extrapolation from galactic data \citep{Hat97,Sha97} or scalings
from extragalactic nova surveys \citep{DVL94,Sha00}: rates based
on extrapolations, 15-24 yr$^{-1}$ or $27\pm 8$ yr$^{-1}$, tend to
be smaller than those based on galactic data, $35\pm 11$ yr$^{-1}$
or $41 \pm 20$ yr$^{-1}$ \citep[a recent review of the Galactic
nova rate can be found in][]{Sha02}.

The optical light curves of classical novae show that there is an
increase in luminosity which corresponds to a decrease of $m_V$
(apparent visual magnitude) of more than 9 magnitudes occurring in
a few days, with a pre-maximum halt, 2 magnitudes before maximum,
in some cases \citep[see][and references therein]{War95}. Nova
light curves are classified according to their speed class,
defined from either $t_2$ or $t_3$, which is the time needed to
decay by 2 or 3 visual magnitudes after maximum. Speed classes
range from very fast ($t_2 < 10 $ days) and fast ($t_2 \sim 11-25$
days) to very slow ($t_2 \sim 151-250$ days) \citep{Pay57}. Some
examples are the fast nova N~Cyg 1992, which had $t_2
\sim 12$ days, the even faster nova N~Her 1991 ($t_2 \sim
2$ days), and the slow nova N~Cas 1993, which had $t_2
\sim 100$ days.

{\it Here I would like to emphasize the essential role played in the
compilation of nova light curves by amateur astronomers and,
particularly, the AAVSO, under the
enthousiast and extremely competent leadership of
the late Janet Mattei, to whom this cataclysmic variable conference
has been dedicated.}

\begin{figure}[t]
\center
\resizebox{0.6\textwidth}{!}{\plotone{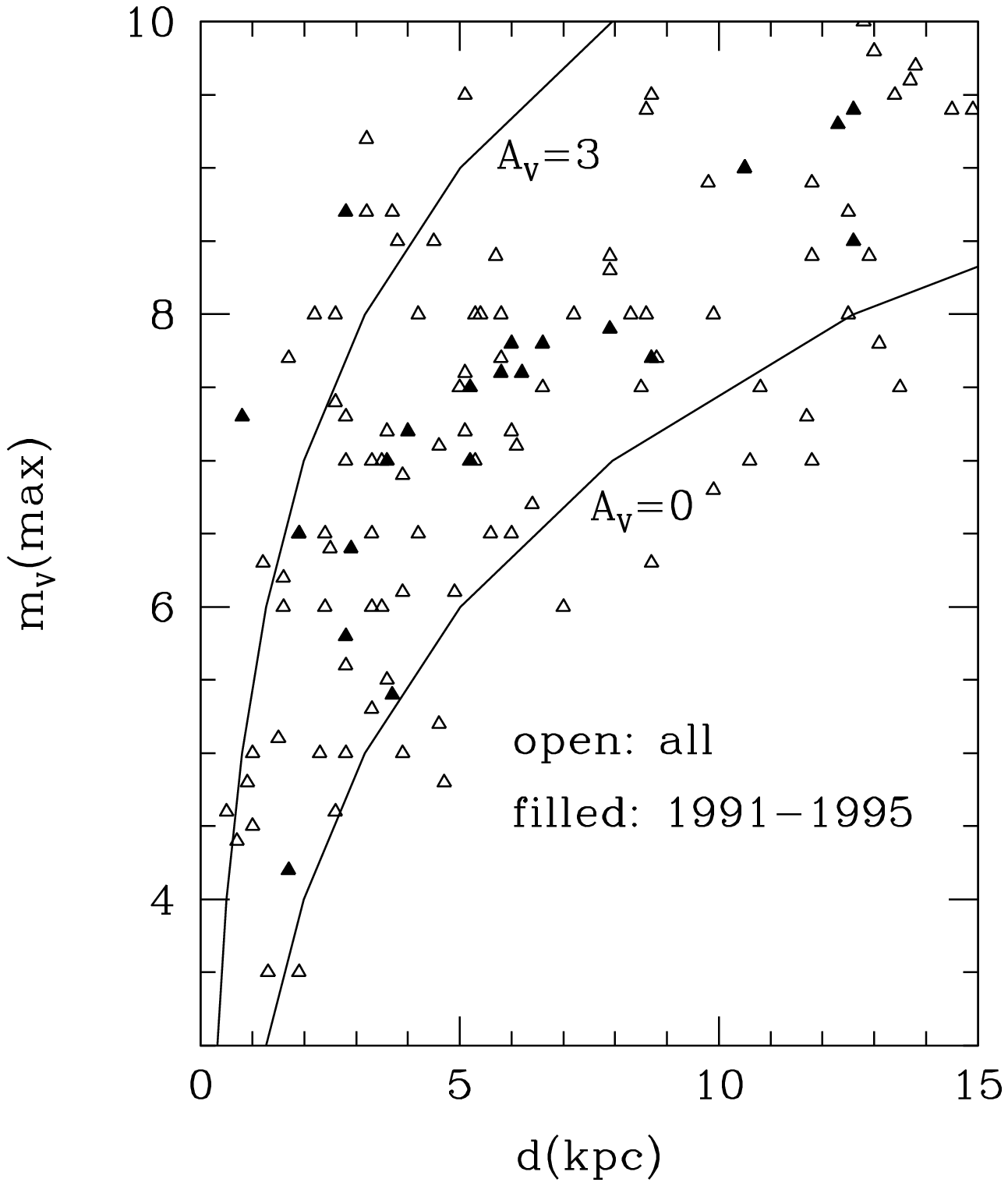}}
\caption{Apparent magnitude at maximum versus distance, for the
sample from \cite{Sha97}, including novae from 1901-1995. The
empirical relation corresponding to visual extinction $A_V=0$ and
$A_V=3$ is shown for illustrative purposes. The MMRD relation
adopted is that of \cite{Sha97}, based mainly on \cite{DVL95}}
\label{fig:mvdist}
\end{figure}

There is a relationship between the absolute magnitude at maximum
$M_V$ and the speed class of novae: brighter novae have shorter
decay times ($t_2$ or $t_3$). The theoretical explanation of this
relationship \citep{her-Liv92} is based on the widely accepted
model of nova explosions, which states that novae reach a maximum
luminosity close to the Eddington luminosity and that they should
eject roughly all their envelope in a time similar to $t_3$.
Therefore, it can be established that $L_{\rm max}$ is an
increasing function of $M_{\rm wd}$ and that $t_3$ is a decreasing
function of $M_{\rm wd}$. From these two relationships an
expression relating $M_V$ at maximum and $t_3$ is deduced. This
empirical relation, valid both in the $V$ and $B$ photometric
bands, is very often used to determine distances to novae, once
visual extinction is known (see figure \ref{fig:mvdist}).
Different calibrations of the maximum magnitude-rate of decline
relationship (MMRD) exist, with that from \cite{DVL95} being the
most usual one \citep[see also][]{Sha97}.

Astronomical satellites like IUE (International Ultraviolet
Explorer), allowed for an extension of the light curves to further
energy ranges besides the optical. It was discovered from IUE
observations that the luminosity in the ultraviolet band increased
when the optical one started to decline: the reason is that the
energy distribution shifts to higher energies, since deeper and
hotter regions of the expanding envelope are seen (the photosphere
recedes as a consequence of the decreasing opacity when
temperature falls below 10$^4$K, the hydrogen recombination
temperature). Infrared observations (when available, i.e. for
novae which form dust) indicate an increase in luminosity once the
ultraviolet luminosity starts to decline, which is interpreted as
the resulting re-radiation (in the infrared) by dust grains of the
ultraviolet energy they have absorbed. In summary, the bolometric
luminosity of classical novae is constant for quite a long period
of time, being the duration of this constant $L_{\rm bol}$ phase
dependent on the remaining envelope mass of the nova. The phase of
constant $L_{\rm bol}$ corresponds to hydrostatic hydrogen burning
in the remaining envelope of the nova, accompanied by continuous
mass-loss, probably due to an optically thick wind \citep{KH94}.
Since the bolometric luminosity deduced from observations is close
to or even larger than the Eddington luminosity, radiation
pressure is probably the main responsible for ejection of nova
envelopes. An additional observational proof of this phase has
come from the observations in the soft X-ray range
\citep{Kra96,Bal98,Ori01}, which directly reveal the remaining
nuclear burning shell on top of the white dwarf after its nova
explosion.

Coming back to the problem of the galactic nova rate, it is worth
reminding that it can not be disentangled from the fact that there
are two distinct nova populations: disk novae, which are in
general fast and bright ($M_V(max) \simeq -8$), and bulge novae,
slower and dimmer ($M_V(max) \simeq -7$). This was first suggested
by \cite{DeV92} and later corroborated by the classification of
novae in two classes according to their early post-outburst
spectra \citep{Wil92} and based on the stronger group of emission
lines they display (either Fe II lines or He and N lines);
FeII-type novae evolve more slowly and have a lower level of
ionization, whereas He/N-type novae have larger expansion
velocities and a higher level of ionization. It has been deduced
that the faster and brighter He/N novae are concentrated closer to
the galactic plane than the slower and dimmer Fe-II ones, which
would preferentially belong to the bulge (\citealt{DVL98}; see
\citealt{DeV02} for a recent review).

In fact, nova spectra are rather complicated and not well
understood from the theoretical point of view. A recent paper by
\cite{Cas04} makes an excellent analysis of IUE archival high
resolution spectra of nova~Cygni 1992. The complex
behavior of the various lines at different epochs after the
outburst requires a complex interpretation. We summarize the
results of \cite{Cas04} here just to show that the complexity of
the observed features leads to an empirical model based on various
mass loss phases, which is not yet well accounted for in the
thermonuclear runaway nova models described below; so observations
are still far ahead from models in this case. In the earliest days
there is an optically thick high velocity wind with decreasing
mass loss rate as time elapses (then the photosphere recedes,
leading to an increasing degree of ionization which manifests in
the strengthening of higher excitation emission lines); emission
comes from successively deeper shells with smaller velocities and
therefore the emission lines get narrower. The absorption lines
superimposed on the emission lines are formed in two different
shells which should have been ejected with some time interval; the
inner shell (ejected later) has larger velocity and catches up the
outer shell. The model in \cite{Cas04} is in agreement with the
optically thick wind models from \cite{KH94}, already addressed by
\cite{Fri66}. Another independent view of the spectral evolution
of Nova~Cyg 1992 (and other objects) can be found in the
recent review by \cite{Sho02}.

A very important result deduced from nova observations (optical,
ultraviolet and infrared) is that their ejecta are very often
enriched in carbon, nitrogen and oxygen, as well as in neon in
many objects (around 1/3 of the total); the global metallicities
in nova ejecta are well above solar metallicites \citep[see][for a
recent review]{Geh98}. This observational fact has been one of the
main drivers of the theoretical models and, of course, should be
explained by them. The general assumption made long ago (see for
instance \citealt{Sta78}; and \citealt{Pri78}) is that some
enrichment of the accreted matter (which is in principle assumed
to be of solar composition) with matter from the underlying white
dwarf core (of the CO, carbon-oxygen, or ONe, oxygen-neon, type)
is necessary, both to power the nova explosion and to explain the
observed enhanced (with respect to solar) abundances.

\section{Scenario}

\subsection{Thermonuclear runaway}

There is a general agreement on the scenario of classical nova
explosions: a low luminosity white dwarf accretes hydrogen-rich
matter in a cataclysmic binary system, as a result of Roche lobe
overflow of its main sequence companion. For accretion rates low
enough, e.g. $\dot{M} \sim 10^{-9}-10^{-10}$ M$_\odot$ yr$^{-1}$,
accreted hydrogen is compressed up to degenerate conditions until
ignition, thus leading to thermonuclear burning without control
(thermonuclear runaway, TNR). Explosive hydrogen burning
synthesizes some $\beta^+$-unstable nuclei of short lifetimes
(e.g. $^{13}$N, $^{14}$O, $^{15}$O, $^{17}$F, with $\tau$ = 862,
102, 176, and 93s respectively) which are transported by
convection to the outer envelope, where they are preserved from
destruction. These decays lead to a huge energy release in the
outer shells which causes the nova outburst, i.e. a visual
luminosity increase accompanied by mass ejection with typical
velocities $10^2-10^3$ km/s.

\begin{figure}[t]
\includegraphics[angle=-90,width=0.8\textwidth]{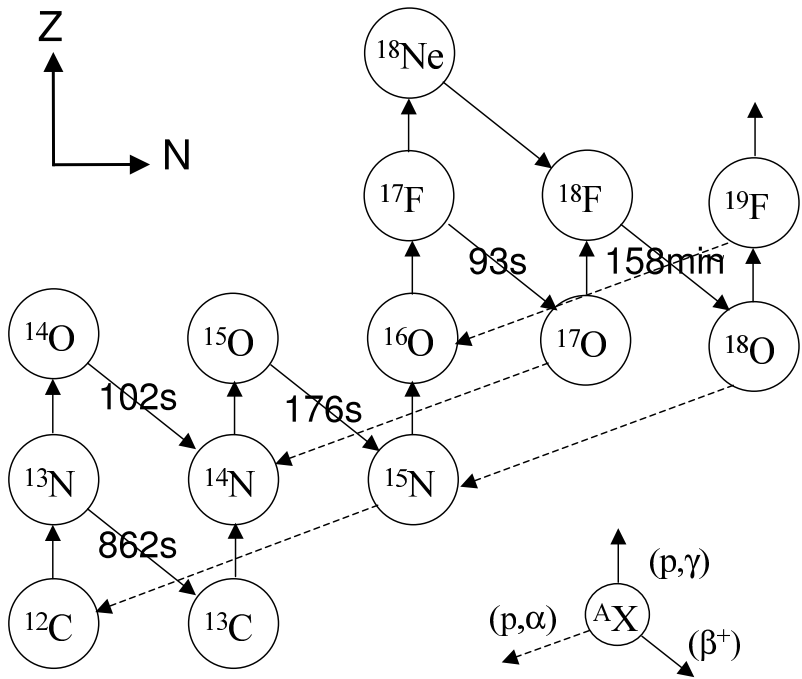}
\caption{Scheme of the carbon-nitrogen-oxygen (CNO) cycle of
hydrogen burning, which operates out of equilibrium in nova
outbursts. The lifetimes of the $\beta^+$-unstable nuclei, which
act as {\it bottlenecks} of the cycle, are displayed}
\label{fig:cno}
\end{figure}

The mechanism for nova explosions is better understood after
evaluating some relevant timescales \citep[see][for a
review]{Sta89}: the accretion timescale, defined as $\tau_{\rm
acc} \sim {\rm M_{acc} / \dot{M}}$ (which is of the order of
$10^4-10^5$ yr, depending on the accretion rate $\dot{M}$ and
accreted mass $M_{\rm acc}$), the nuclear timescale $\tau_{\rm
nuc} \sim C_{\rm p} T / \epsilon_{\rm nuc}$ (which is as small as
a few seconds at peak burning; $C_{\rm p}$ is the specific heat
and $\epsilon_{\rm nuc}$ the rate of nuclear energy generation),
and the dynamical timescale ($\tau_{\rm dyn} \sim H_{\rm p} /
c_{\rm s} \sim (1/g) \sqrt{P/\rho}$; $H_{\rm p}$ is the pressure
scale height and $c_{\rm s}$ the local sound speed). During the
accretion phase, $\tau_{\rm acc} \le \tau_{\rm nuc}$, accretion
proceeds and the envelope mass increases. When degenerate ignition
conditions are reached, degeneracy prevents envelope expansion and
the TNR occurs. As temperature increases, degeneracy would be
lifted (since $T$ would become larger than $T_{\rm Fermi}$) and
expansion would turn-off the explosion, but this is not the case
because $\tau_{\rm nuc} \ll \tau_{\rm dyn}$ (specially if the
envelope is enriched in CNO elements, thus enhancing the
contribution of the CNO cycle to hydrogen burning). Therefore,
since the envelope can not readjust itself through expansion,
temperature and nuclear energy generation rate continue to
increase without control. The value of the nuclear timescale is
crucial for the development of the TNR and its final fate. In fact
there are mainly two types of nuclear timescales: those related to
$\beta^+$-decays, $\tau_{\beta^+}$, and those related to proton
capture reactions, $\tau_{(\rm p,\gamma)}$. In the early evolution
towards the TNR, $\tau_{\beta^+} < \tau_{(\rm p,\gamma)}$ and the
CNO cycle operates in equilibrium. But as temperature increases up
to $\sim 10^8$ K, the reverse situation is true ($\tau_{\beta^+}
\ga \tau_{(\rm p,\gamma)}$), and thus the CNO cycle is
$\beta$-limited (see figure \ref{fig:cno}). In addition, since the
large energetic output produced by nuclear reactions can not be
evacuated only by radiation, convection sets in and transports the
$\beta^+$-unstable nuclei to the outer cooler regions where they
are preserved from destruction and where they will decay later on
($\tau_{\rm conv} \la \tau_{\beta^+}$), leading to envelope
expansion, luminosity increase and mass ejection if the attained
velocities are larger than escape velocity. Another important
effect of convection is that it transports fresh unburned material
to the burning shell. In summary, non-equilibrium burning occurs
and the resulting nucleosynthesis will be far from that of
hydrostatic hydrogen burning.

\subsection{Underlying white dwarf: CO or ONe}

As mentioned above, mixing between the accreted envelope (with
solar composition) and the underlying white dwarf is a necessary
condition both to power the explosion and to interpret the large
over solar metallicities observed in nova ejecta. There have been
many suggested mechanisms to explain this process, either
occurring prior or during the thermonuclear runaway, but none of
them is complete satisfactory up to now \citep[see][for an
extensive review]{Liv94}. One example is diffusion induced
convection, first discussed by \cite{PK84} and \cite{KP85}, which
can explain moderate enrichments but has difficulties to account
for some large enrichments observed \citep{PK97}. Other
possibilities are shear mixing, convection induced shear mixing
and convective overshooting induced flame propagation. Recent
efforts with multidimensional codes have not yet succeeded
completely in reproducing the necessary mixing needed to power the
explosion (see a recent review in \cite{GL02} and a recent
multidimensional result in \cite{Ale04}).

It is clear that once a mechanism for mixing is adopted, even if
not well physically founded yet, the following step is to take
some composition of the underlying white dwarf core to mix with
the solar (in principle) composition of accreted matter; core
composition will play a crucial role in the subsequent evolution.
It is important to distinguish between novae occurring on CO and
ONe white dwarfs: stars originally less massive than $\sim 10-12$
M$_\odot$ end their life as white dwarfs, made of helium, carbon
and oxygen or oxygen and neon. The exact mass interval leading a
particular white dwarf type is not completely well determined,
since it depends on details of stellar evolution and, especially,
on the single or binary nature of the progenitor. The most common
white dwarf case is CO, whereas for massive progenitors with
masses around 10 M$_\odot$, non-degenerate carbon ignition leads
to the formation of a degenerate core mainly made of oxygen and
neon, with traces of magnesium and sodium. These cores were
thought to be made of oxygen, neon and magnesium (the so-called
ONeMg white dwarfs) some years ago, when parametrized calculations
of hydrostatic carbon burning were adopted \citep{AT69}  before
self-consistent models of AGB (asymptotic giant branch stars)
following the thermally pulsing phase were available
\citep{Dom93,Rit96}. The minimum mass at the zero age main
sequence leading to extensive carbon burning is $\sim 9.3$
M$_\odot$ and the resulting ONe mass is $\sim 1.1$ M$_\odot$, if
the evolution occurs in a binary system \citep{Gil03}. It is
important to notice that the ONe white dwarf has a thick {\it CO
buffer} on top of its ONe core, which would prevent the mixing of
accreted matter with the underlying ONe core (at least until a
number of previous outbursts would have eroded that buffer). Then,
strange explosions could occur where there would be, for instance,
a lack of neon in the ejecta of a nova occurring on top of an ONe
core \citep[see][for details]{Jos03}.

\section{Relevance of nucleosynthesis in novae}

The main goal of theoretical models of nucleosynthesis in novae is
to reproduce the observed abundances in novae ejecta. Although
both from the observational and theoretical side some
uncertainties exist, a rather good fit is obtained in many cases
(e.g. see table 5 in \citealt{JH98}).

\begin{figure}[t]
\plottwo{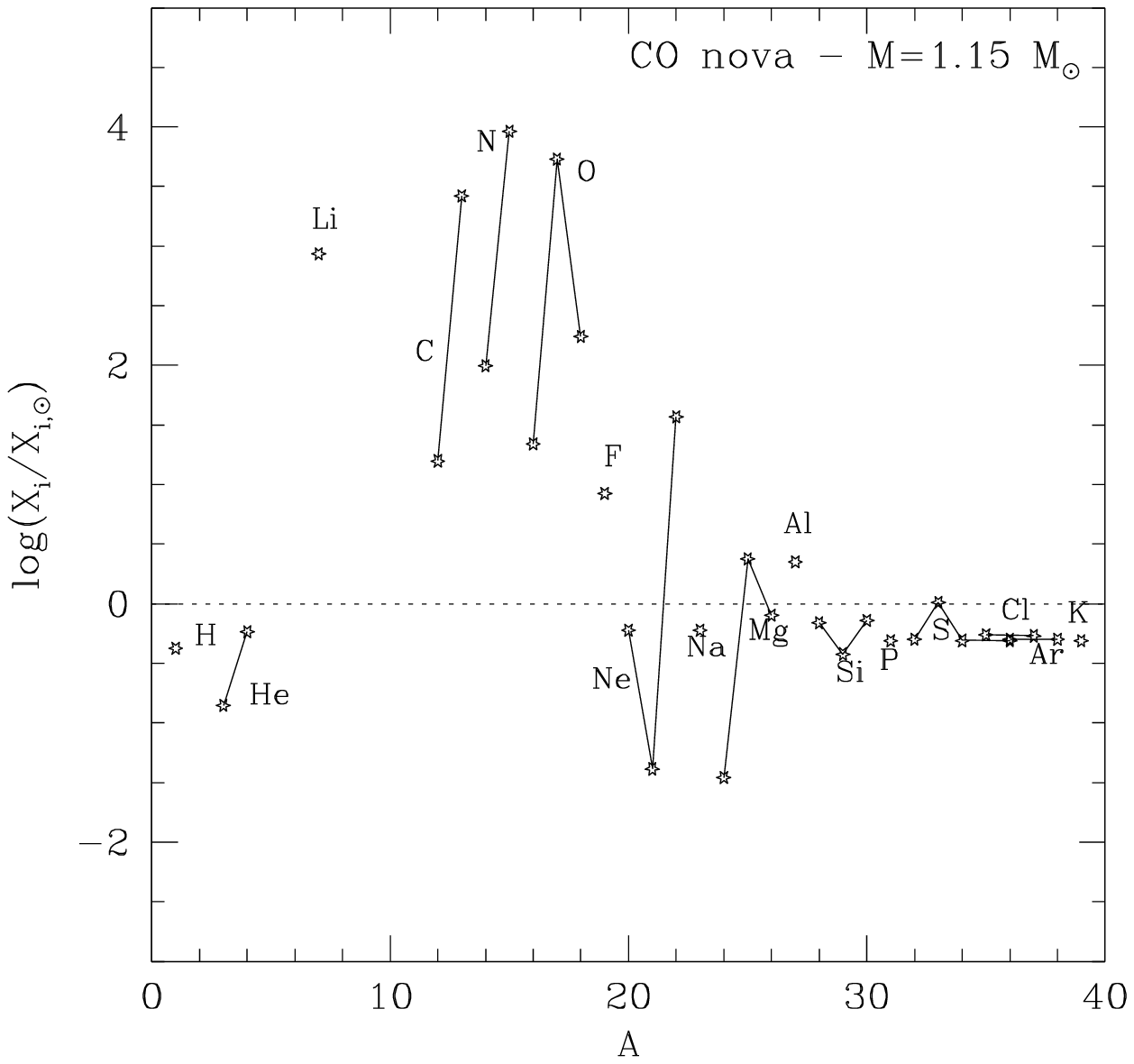}{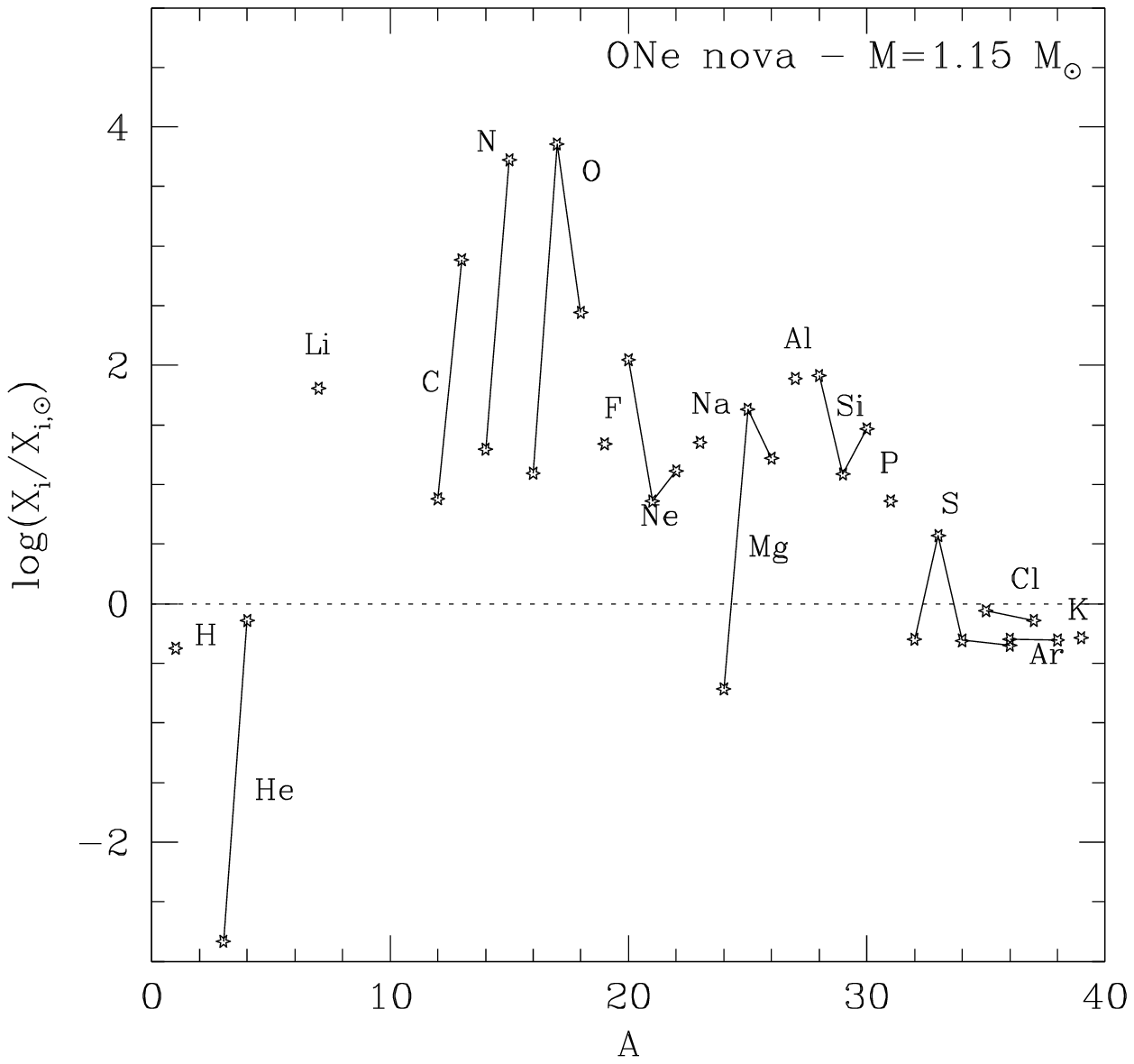}
\caption{Logarithmic overproduction factors with respect to solar
abundances, versus mass number. Left panel: CO nova of 1.15
M$_\odot$. Right panel: ONe nova of 1.15 M$_\odot$.}
\label{fig:abundances1}
\end{figure}

\begin{figure}[t]
\center
\resizebox{0.8\textwidth}{!}{\plotone{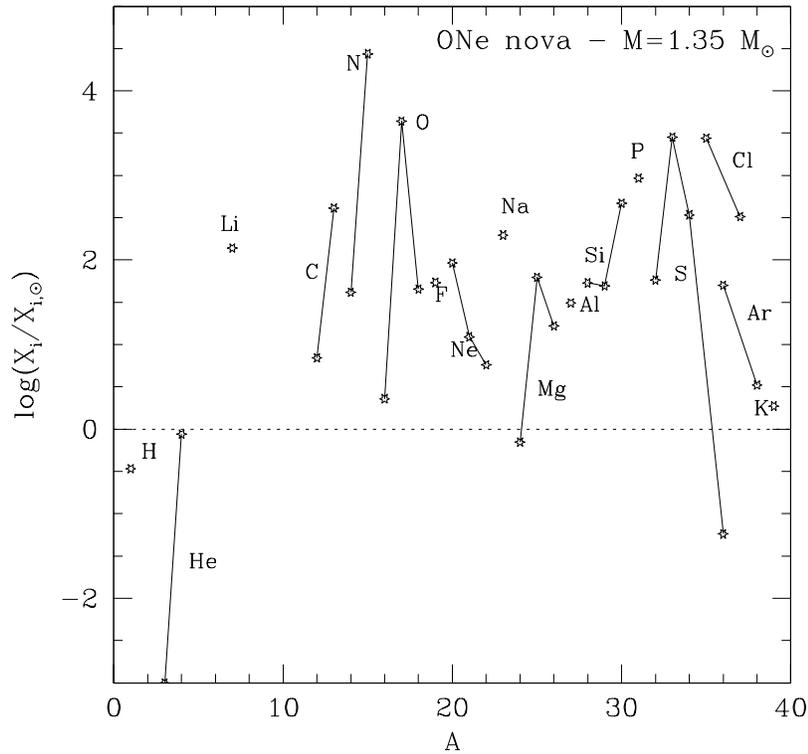}}
\caption{Same as figure \ref{fig:abundances1} but for an ONe nova
of 1.35 M$_\odot$} \label{fig:abundances2}
\end{figure}

Novae are not important contributors to the abundances observed in
the interstellar medium, in contrast with supernovae. However,
they can contribute to Galactic abundances of some particular
isotopes, whenever the overproduction factors with respect to
solar abundances are larger than around $10^3$ (see
\citealt{JH98}; and \citealt{Geh98}). In figure
\ref{fig:abundances1} the overproduction factors relative to solar
versus mass number are shown for two typical novae: a CO and an
ONe novae, with mass 1.15 M$_\odot$, 50\% of mixing with core
material and accretion rate $2\times 10^{-10}$ M$_\odot$
yr$^{-1}$. In this figure some general features are
distinguishable: both in CO and in ONe novae, the largest yields
correspond to elements of the CNO group, whether in CO novae
$^7$Li is also largely overproduced. In the case of more massive
ONe novae (see figure \ref{fig:abundances2} corresponding to mass
1.35 M$_\odot$), intermediate-mass elements such as Ne, Na, Mg, S,
Cl are also overproduced \citep{JH98}.

The large overproduction of $^7$Li in CO novae is an important
result, since the origin of Galactic lithium is still not well
understood \citep{Her96}. It is widely accepted that there is some
primordial lithium produced during the big bang and that
spallation reactions by cosmic rays in the interstellar medium or
in stellar flares also produce it. But some extra stellar source
of $^7$Li (without generating $^6$Li) has to be invoked to explain
the steep rise of the observed lithium abundance between the
formation of the solar system and the present time \citep{Rom99}.
Concerning the CNO group isotopes, novae largely overproduce
$^{13}$C, $^{15}$N and $^{17}$O: Galactic $^{17}$O is most
probably almost entirely of novae origin \citep{JH98}, whereas for
$^{13}$C and $^{15}$N other sources are probably required.

Another important piece of information relative to nova nucleosynthesis
comes from the measurement of isotopic ratios in presolar grains. Five
SiC (silicon carbide) and one graphite grain isolated from the Murchison
carbonaceous meteorite have been discovered with low $^{12}$C/$^{13}$C
and $^{14}$N/$^{15}$N ratios, large excesses in $^{30}$Si and high
$^{26}$Al/$^{27}$Al ratios. These isotopic signatures are close to those
predicted theoretically for ONe ejecta and can not be matched by any other
stellar source \citep{Ama01}. Recent more detailed theoretical
estimates will help to properly identify nova grains in primitive
meteorites with nova origin \citep{Jos04}. The formation of grains in
novae is fully justified, since infrared observations clearly indicate that
some novae form dust \citep{Geh98}.

\section{High energy emission from novae}

\subsection{Gamma-rays}

\begin{figure}[h]
\plottwo{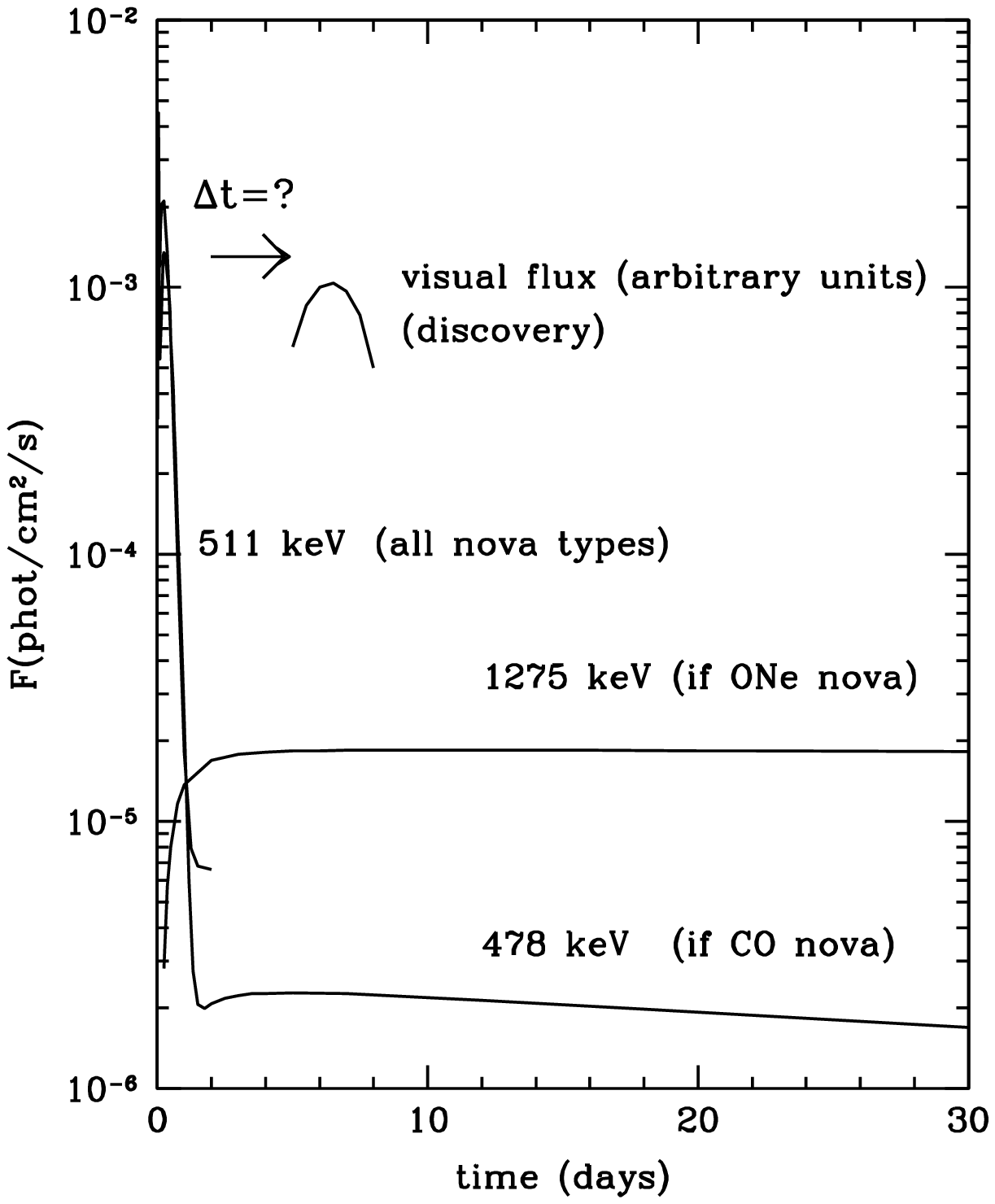}{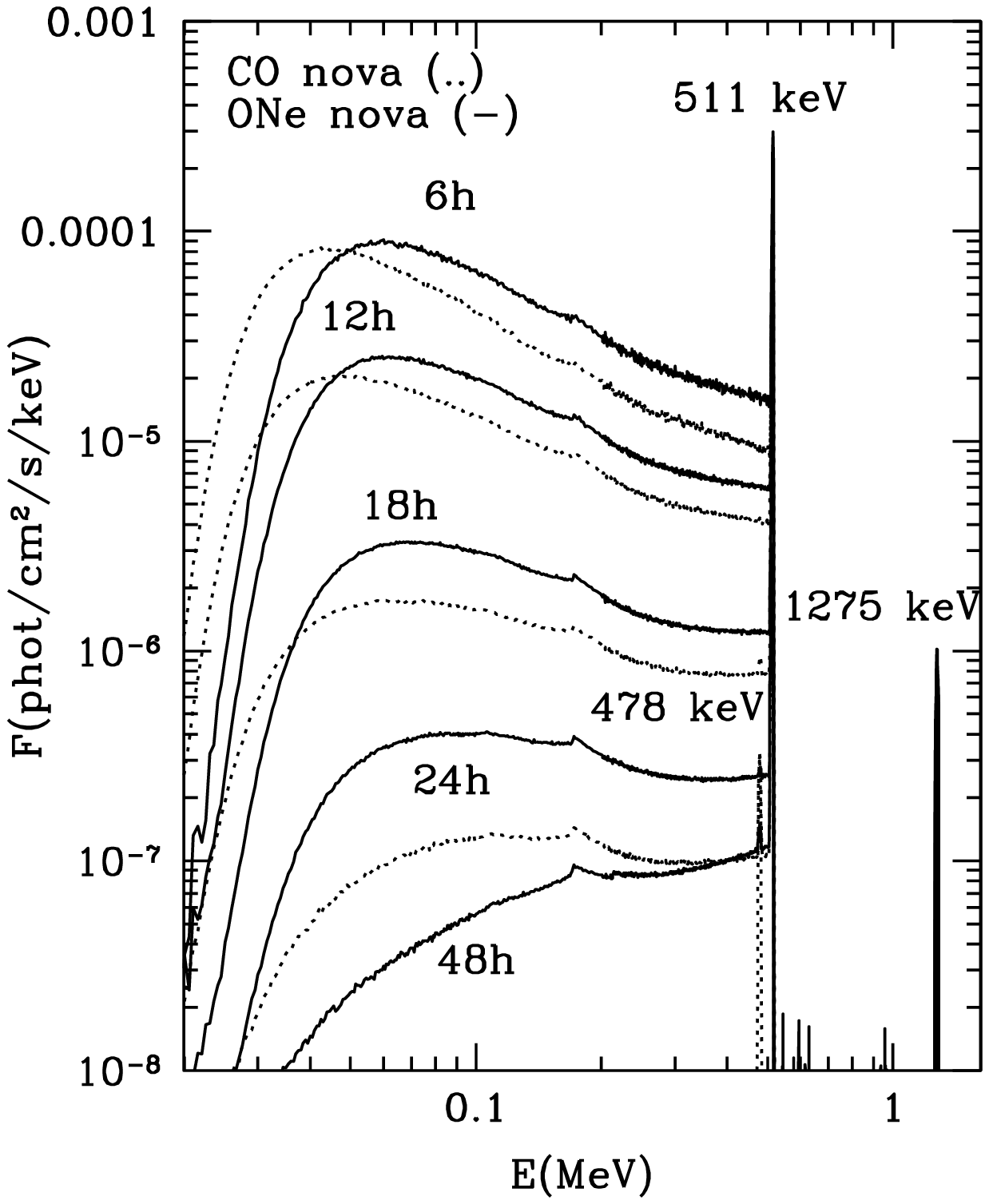}
\caption{Left panel: Gamma-ray light curves of the three possible
lines as compared with visual light curve (the latter in arbitrary
units). Right panel:Early temporal evolution of the gamma-ray
spectra of a CO (dotted) and an ONe (solid) nova, at a distance of
1 kpc.} \label{fig:gamma}
\end{figure}

An important property of novae ejecta is the presence of
radioactive nuclei (the role of novae as potential $\gamma$-ray
emitters was mentioned long ago by \cite{CH74,Cla81,LC87}). In
addition to the very short-lived isotopes responsible for the nova
explosion, other longer-lived nuclei are synthesized which have
some relevance to the radioactivity of the Galaxy and to the
$\gamma$-ray emission of individual novae.

Short-lived nuclei $^{13}$N and $^{18}$F ($\tau$=862s and 158 min)
are produced in similar quantities in all nova types, whereas
$^{7}$Be ($\tau$=77days) is mainly produced in CO novae and
$^{22}$Na ($\tau$=3.75 yrs) and $^{26}$Al ($\tau=10^6$ yrs) are
produced in appreciable amounts only in ONe novae. The reason is
that in nova explosions the temperatures reached (around $2-3
\times 10^8$ K) are not high enough to break the CNO cycle;
therefore, only if some seed nuclei (like $^{20}$Ne, $^{23}$Na,
$^{24,25}$Mg) are present in the envelope material, can the
NeNa-MgAl cycles operate and synthesize those radioactive nuclei
(and other intermediate-mass isotopes). As CO white dwarfs are
devoid of these nuclei, it is almost impossible for them to
produce large amounts of radioactive $^{22}$Na and $^{26}$Al.

Radioactive nuclei ejected by novae play a role in the
radioactivity of the Galaxy which depends on their lifetimes. The
short-lived nuclei (i.e., $^{13}$N and $^{18}$F) produce an
intense burst of $\gamma$-ray emission, with duration of some
hours, which is emitted before the nova visual maximum (see figure
\ref{fig:gamma} and \cite{Gom98,Her99} for details). This emission
is related to positron annihilation, which consists of a line at
511 keV and a continuum at energies between 20 and 511 keV,
related to the positronium continuum plus the comptonization of
the photons emitted in the line. In figure \ref{fig:gamma} the
spectral evolution of a CO and an ONe nova at different epochs
after peak temperature is shown. The emission related to
medium-lived nuclei, $^{7}$Be and $^{22}$Na, appears later and is
different in CO and ONe novae, because of their different
nucleosynthesis: CO novae display a line at 478 keV related to
$^{7}$Be decay, whereas ONe novae show a line at 1275 keV related
to $^{22}$Na decay.

The long-lived isotope $^{26}$Al is also produced by novae. The
Galactic $\gamma$-ray emission observed at 1809 keV (\cite{Mah84}
with the HEAO 3 satellite; \cite{Die95} with the CGRO/COMPTEL)
corresponds to the decay of $^{26}$Al. Its distribution seems to
correspond better to that of a young population and the
contribution of novae is not the dominant one
\citep[see][]{PD96,JHC97}.

In summary, classical novae explosions produce $\gamma$-rays,
being the signature of CO and ONe novae different. The
detectability distances for the lines at 478 and 1275 keV with the
INTEGRAL/SPI instrument ranges between 0.2 and 0.5 kpc
\citep{HJ04}; the width of the lines ($\sim$ 7 keV for the 478 keV
line and $\sim$ 20 keV for the 1275 keV line), which is non
negligible and largely degrades the sensitivity of the instrument,
has been taken into account to compute these distances. The
continuum and the 511 keV line are the most intense emissions, but
their appearance before visual maximum and their very short
duration requires ``a posteriori" analyses, with large
field-of-view instruments monitoring the whole sky at the
appropriate energies (some hundred keVs). With future instruments
of these characteristics, novae would be detectable more easily in
$\gamma$-rays than visually, because of the lack of extinction. So
future instrumentation in the $\gamma$ and hard X-ray domain will
give crucial insights on the nova theory allowing for a direct
confirmation of the nucleosynthesis in these explosions, but also
providing unique information about the Galactic distribution of
novae and their rates.

\subsection{X-rays}

X-rays provide an important piece of information about nova
outbursts, specially concerning their turnoff. The {\it EXOSAT}
satellite detected the nova GQ~Mus (N Mus 1983) as a soft
X-ray emitter (in the interval 0.04-2 keV), 460 days after optical
maximum \citep{Oge84}. {\it ROSAT} detected again that source
(0.1-2.4 keV), even 9 years after the explosion \citep{Oge93}.
Nova~Cyg 1992 was also detected by {\it ROSAT} as a
bright soft X-ray source, but its emission lasted only for 18
months \citep{Kra96}. The soft X-ray emission is interpreted as
the photospheric emission of the hot white dwarf, hosting a
remaining hydrogen-burning shell, which becomes visible when the
expanding envelope is transparent enough. The luminosity in soft
X-rays is close to $L_{\rm Edd}$, thus indicating again the
constancy of $L_{\rm bol}$ and the hardening of the spectra. The
turn-off times of novae deduced from soft X-ray and ultraviolet
observations, range between 1 and 5 yr \citep{GR98}, except for
N~Mus 1983 (9-10 yr). This is much shorter than expected
from the nuclear burning timescale of the generally accepted mass
of the remaining envelope (see Sala \& Hernanz, this volume);
therefore, some extra and unknown mechanism should remove mass
after or during the nova outburst (see as well the papers by Orio,
Balman and Hernanz in this same volume, for recent results from
X-ray observations of novae -- and nova remnants -- with
XMM-Newton and Chandra).

In addition to soft X-rays, novae can also emit harder X-rays, as
observed with {\it ROSAT} (\citealt{Llo92}; \citealt*{Ori01};
\citealt{Kra02}) and more recently with XMM-Newton and Chandra. We
would just like to mention that observations of Nova~Oph
1998 with {\it XMM-Newton}, have clearly revealed that accretion
had been reestablished as soon as 2.7 years after explosion, which
indeed is a very short recovery time \citep{HS02}, perhaps
indicating that accretion disks (or in general {\it accretion
streams}, if the system is magnetic) are not completely destroyed
by the explosion.

\section{Discussion}
A number of open problems concerning classical novae that still
remain unsolved were outlined in the discussion session of the
conference:

\begin{itemize}
\item Accretion rates: variability and relationship with the
hibernation scenario.

\item Ejected mass: can theory be reconciled with observations?

\item High metallicity ejecta: how to explain it? How does the
required mixing occur?

\item How long is the decline from maximum bolometric luminosity
to pre-outburst luminosity?

\item Super Eddington luminosities in white dwarfs.

\item Long-term evolution of white dwarfs in novae: does the white
dwarf grow in mass? Possible scenario of type Ia supernovae.
\end{itemize}

These topics have been since long in the list of open and unsolved
questions and they remain there, in spite of the observational and
theoretical efforts of the last years. Hopefully some of them will
be better understood when the next meeting takes place.

\acknowledgments{Financial support from the MCYT through the project
AYA2001-2360 and by the EU FEDER funds is acknowledged.}

\end{document}